# Crystallization kinetics of equimolar liquid crystalline mixture and its components

Aleksandra Deptuch[1,*], Anna Paliga[2], Anna Drzewicz[1], Marcin Piwowarczyk[1], Magdalena Urbańska[3], Ewa Juszyńska-Gałązka[1,4]

[1] Institute of Nuclear Physics Polish Academy of Sciences, Radzikowskiego 152, PL-31342 Kraków, Poland

[2] Faculty of Physics and Applied Computer Science, AGH University of Kraków, Reymonta 19, PL-30059 Kraków, Poland

[3] Institute of Chemistry, Military University of Technology, Kaliskiego 2, PL-00908 Warsaw, Poland

[4] Research Center for Thermal and Entropic Science, Graduate School of Science, Osaka University, 560-0043 Osaka, Japan

*corresponding author, aleksandra.deptuch@ifj.edu.pl

**Abstract**

The new equimolar mixture comprises liquid crystalline compounds MHPOBC and partially fluorinated 3F2HPhF6. The phase sequence of the mixture is determined by differential scanning calorimetry, polarizing optical microscopy, X-ray diffraction, and broadband dielectric spectroscopy. The enantiotropic smectic A*, C*, and $C_A$* phases are observed for the mixture. Only partial crystallization of the mixture is observed during cooling at 2-40 K/min and the remaining smectic $C_A$* phase undergoes vitrification. In contrast, the crystallization of the pure components is complete or almost complete for the same range of cooling rates. The kinetics of the non-isothermal and isothermal crystallization of the mixture and pure components are investigated by differential scanning calorimetry. The non-isothermal data are analyzed by the isoconversional method, while the isothermal data are analyzed using the Avrami model. Typically, the nucleation-controlled crystallization kinetics are observed.

**1. Introduction**

The smectic liquid crystals (smectic LCs), possessing the lamellar order, can show the switching in the electric field if particular conditions are fulfilled: the molecules are chiral, the tilted smectic phase is present (SmC* or its sub-phases), the component of the molecular dipole moment is non-zero in a direction perpendicular to the plane of tilt, and the helical twist of the molecular tilt along the layer normal is removed [1,2]. Smectogenic compounds with such properties can be used in LC displays [3]. To control the temperature range of the tilted smectic phase, which defines the operating conditions of the display, numerous mixtures have been investigated, e.g. [4-12].

The 4-[(1-methylheptyloxy)carbonyl]phenyl 4′-octyloxy-4-biphenylcarboxylate compound, abbreviated as MHPOBC, is the first LC where the tristable switching in the electric field was detected, indicating the existence of the antiferroelectric smectic $C_A$* phase (Sm$C_A$*) [13]. MHPOBC was tested as a component of mixtures with other LCs [5-8] and as a nanoparticle matrix in composites [14-18]. The phase sequence on cooling given in [19] is Iso (421 K) SmA* (395 K) Sm$C_α$* (394.1 K) SmC* (392.4 K) Sm$C_γ$* (391.6 K) Sm$C_A$* (337.7 K) Sm$X_A$* (304 K) Cr, where Iso is the isotropic liquid, Sm denotes various smectic phases, and Cr is the crystal phase. In the SmA* phase, the average tilt angle of molecules is zero, and this phase has paraelectric properties. The three next phases, present in a narrow temperature range, are interpreted in various ways in the literature. The resonant X-ray diffraction study of (S)-MHPOBC films, with a small admixture (up to 5% molar percentage) of the selenophene PB237, leads to the identification of these phases as ferroelectric SmC*, four-layer $SmC_{FI2}$*, and three-layer $SmC_{FI1}$* in order of decreasing temperature [20]. However, it is underlined in [20] that the PB237 dopant, necessary for the resonant X-ray experiment, might alter the structure of the smectic phases. The presence of optical impurities also changes the phase sequence. According to [21], in pure (S)-MHPOBC, these are $SmC_\alpha$*, $SmC_{FI2}$*, and $SmC_{FI1}$*. Meanwhile, for (S)-MHPOBC with a small admixture of (R)-MHPOBC, the $SmC_{FI2}$* phase is replaced by SmC*, and $SmC_\alpha$* is interpreted as a phase with a short helical pitch both for optically pure (S)-MHPOBC and (S)-MHPOBC with (R)-MHPOBC admixture up to 12% [21]. The sequence $SmC_\alpha$*, SmC*, $SmC_{FI1}$* is assumed in a recent paper [22]. The $SmC_A$* phase has a wide temperature range. After supercooling, it transitions to the monotropic $SmX_A$* phase by the appearance of the hexatic bond-orientational order within layers [23]. The $SmX_A$* phase is either $SmI_A$* or $SmF_A$* (in [19] denoted as $SmI_A$*), which are difficult to distinguish, as the only difference in their structures is the direction of the molecular tilt in respect to the local hexagonal order within layers. This usually cannot be inferred from powder X-ray diffraction patterns [23]. We did not find any XRD study of the oriented MHPOBC sample in the literature, which would distinguish $SmI_A$* and $SmF_A$*, therefore we prefer to use the $SmX_A$* notation for the hexatic phase.

The second investigated compound, (S)-4'-(1-methylheptyloxycarbonyl)biphenyl-4-yl 4-[2-(2,2,3,3,4,4,4-heptafluorobutoxy)etyl-1-oxy]-2-fluorobenzoate, is abbreviated as 3F2HPhF6 and exhibits only two smectic phases, with a sequence Iso (377.2 K) SmC* (364.6 K) $SmC_A$* (307 K) Cr1 (299.5 K) Cr2 on cooling [24]. The $SmC_A$* phase of 3F2HPhF6 is orthoconic, meaning the molecules' tilt angle reaches almost 45° [24]. This property improves the quality of the dark state in an LC display [25,26]. The results for one LC mixture containing 3F2HPhF6 were published [4]. The polymorphism of the MHPOBC/3F2HPhF6 system has not been investigated yet.

The paper aims to investigate the crystallization process of MHPOBC, 3F2HPhF6, and their equimolar mixture by differential scanning calorimetry (DSC) for various cooling rates and in isothermal conditions at different temperatures. This is the part of the broader study of the crystallization kinetics in antiferroelectric smectic liquid crystals, which has been focused on pure compounds up to now [27-30]. The sequence of the smectic phases in the mixture is also determined by DSC and polarizing optical microscopy (POM). X-ray diffraction (XRD) is used to measure the smectic layer spacing in the mixture as a function of temperature in comparison to values for pure components. Broadband dielectric spectroscopy (BDS) facilitates the identification of the smectic phases based on the relaxation processes observed in the mixture.

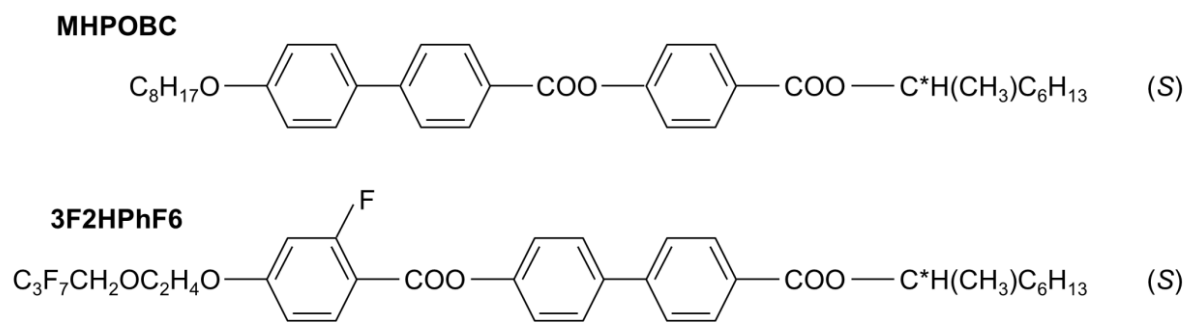

Figure 1. Molecular structures of compounds used in this study.

**2. Experimental details**

The MHPOBC [21] and 3F2HPhF6 [4] compounds, both S-enantiomers, were synthesized at the Institute of Chemistry, Military University of Technology in Warsaw. The components were mixed in the molar fraction 0.5002(7) : 0.4998(6), respectively. The mixture was prepared by the dissolution of both components in acetone in 313 K and the evaporation of the solvent. Then, the precipitate was heated to 433 K to ensure proper mixing in the isotropic liquid phase. The obtained mixture is referred to as MIX2HF6.

The DSC measurements were performed for the 2.180 mg, 2.260 mg, and 6.000 mg samples of MHPOBC, 3F2HPhF6, and MIX2HF6, respectively, in the aluminum pans for the cooling/heating rates in the 2-40 K/min range using the DSC 2500 TA Instruments (New Castle, DE, US) calorimeter. Additionally, the modulated-temperature DSC (MTDSC) measurement for MIX2HF6 was performed during heating at 3 K/min and modulation of 1 K per 60 s, after quenching the sample to 153 K. The DSC thermograms were analyzed using the TRIOS and OriginPro programs.

The samples for POM observations were prepared as a film between two glass slides without aligning layers. The textures were collected using the Leica (Wetzlar, Germany) DM2700 P polarizing microscope with the Linkam (Waterfield, UK) temperature stage. The phase transition temperatures were determined during cooling and heating at 5 K/min. For MIX2HF6, the additional POM observations for the 30 K/min rate were performed in a broader temperature range. The data analysis was done in the TOApy program [31].

The X-ray diffraction measurements with CuKα radiation were performed in the Bragg-Brentano geometry for flat samples inserted into a holder with the 13 mm × 10 mm area and 0.2 mm depth. The diffraction patterns were registered on cooling from the isotropic liquid, using the X'Pert PRO PANalytical (Malvern, UK) diffractometer with the TTK-450 Anton Paar (Graz, Austria) temperature chamber. The fluorophlogopite mica $Mg_3K[AlF_2O(SiO_3)_3]$ [32,33], which is the NIST Standard Reference Material 675 for low 2θ angles [34], purchased from Merck (Darmstadt, Germany), was used for calibration of the diffraction peak positions. The data analysis was done using the FullProf package [35] and OriginPro.

The BDS measurements were conducted for the samples with a thickness of 80 μm, placed between two gold electrodes with polytetrafluoroethylene spacers preventing shortcut. The dielectric spectra were collected using the Novocontrol (Frankfurt, Germany) impedance spectrometer for frequencies between 0.1 Hz and 10 MHz, on slow cooling for all samples and upon heating after slow and fast cooling for MIX2HF6. The data analysis was done in OriginPro.

## 3. Results and discussion
### 3.1. Phase sequence

The DSC results for MHPOBC, 3F2HPhF6, and MIX2HF6 are presented in Figure 2, 3, and 4, respectively. The POM results are shown in Figures S1-S4 in the Supplementary Materials. The DSC curves of MHPOBC collected during cooling (Figure 2a,b) reveal the phase sequence Iso (418.2 K) SmA* (388.3 K) SmC* (386.6 K) SmC$_A$* (337.3 K) SmX$_A$* (305.3 K) Cr3, where the phase transition temperatures are extrapolated to the zero cooling rate. For the cooling rates 25-40 K/min, the melt crystallization is completed during subsequent heating (Figure 2c,d), and the onset temperature of the cold crystallization is 290-293 K. On further heating, another crystal phase or phases appear. For the 6-40 K/min heating rates, there is the Cr3 → Cr2 transition at 319-329 K, which is followed by the Cr2 → Cr1 transition at 336-338 K. For the 2-5 K/min heating rates, there is the direct Cr3 → Cr1 transition at 304-305 K. The phase transition temperatures of MHPOBC, extrapolated to the zero heating rate, are Cr3 (317.1 K) Cr1 (351.6 K) SmC$_A$* (385.0 K) SmC* (389.3 K) SmA* (416.6 K) Iso. All transition temperatures are the onset temperatures $T_o$ of anomalies in the DSC thermograms [36], except the SmC* → SmA* transition, where the peak temperature $T_p$ was used. Under the polarizing microscope, the MHPOBC sample (Figure S1) shows mainly a fan-shaped texture in SmA* and a broken fan-shaped texture in SmC*, SmC$_A$* [22,23]. In some parts, the homeotropic texture is visible [23]. The areas oriented homeotropically are usually dark, except the 340-360 K region, where the increase in luminance is observed, with a peak at 349 K. It is explained by the helix inversion, which occurs in this temperature range [37]. SmC* or its sub-phases are not visible in the plot of luminance vs. temperature; the texture collected in 394 K is attributed to them based on visual inspection. In the SmX$_A$* phase, the texture is still a broken fan-shaped type, but looks more closely to that of the SmA* phase. The decrease of the luminance below 305 K indicates crystallization. During heating, the transition between the crystal phases Cr2 and Cr1 shows a broad valley in the luminance in the 317-342 K range. The Cr1 phase melts at 358 K.

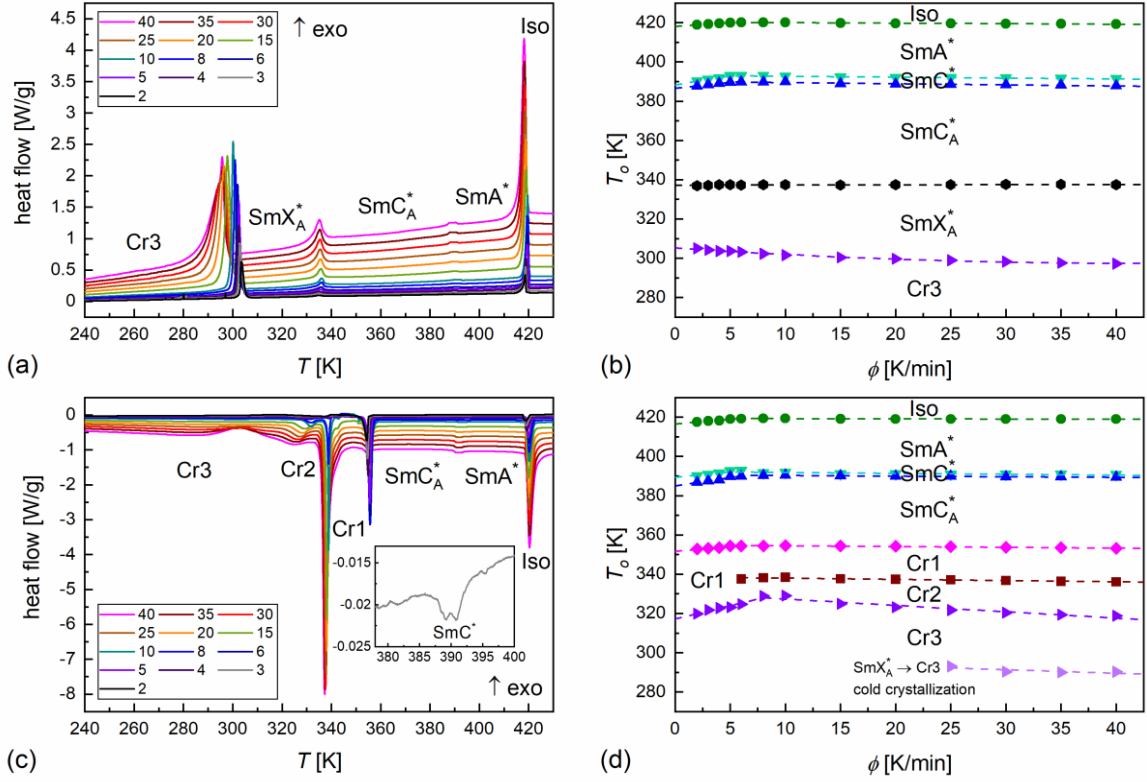

Figure 2. DSC thermograms of MHPOBC collected on cooling at 2-40 K/min rates (a), onset transition temperatures as a function of the cooling rate (b), DSC thermograms collected on heating (c), and onset transition temperatures as a function of the heating rate (d). The inset in (c) shows the DSC thermogram collected during heating at 3 K/min around the SmC$_A$*/SmC*/SmA* transition.

The phase transitions of 3F2HPhF6 observed in the DSC thermograms during cooling (Figure 3a,b), extrapolated to zero cooling rate, are as follows: Iso (377.6 K) SmC* (365.1 K) SmC$_A$* (311.2 K) Cr1. For 2-5 K/min, the sample crystallizes in the phase denoted as Cr1. For 6-10 K/min, a mixture of two crystal phases, Cr1 and Cr2, is formed. For 15-40 K/min, the crystallization to the Cr2 phase is observed. Upon subsequent heating (Figure 3c,d), the cold crystallization of Cr2 from the remaining SmC$_A$* phase occurs for 30-40 K/min, with the onset temperature of 291-292 K. The Cr2 → Cr1 transition is observed at 320-321 K upon heating at 8-40 K/min (for 6 K/min, this anomaly was not visible due to the small amount of Cr2 formed during cooling). The phase transition temperatures extrapolated to zero heating rate are: Cr1 (325.9 K) SmC$_A$* (369.2 K) SmC* (376.9 K) Iso. Under the polarizing microscope, 3F2HPhF6 shows the broken fan-shaped textures in the SmC* and SmC$_A$* phases (Figure S2). The SmC* → SmC$_A$* transition has a weak impact on luminance during cooling and a significant impact during heating. The SmC$_A$* → Cr1 transition occurs gradually below 310 K with visible growing crystallites, and the melting of Cr1 during heating is observed at 328 K.

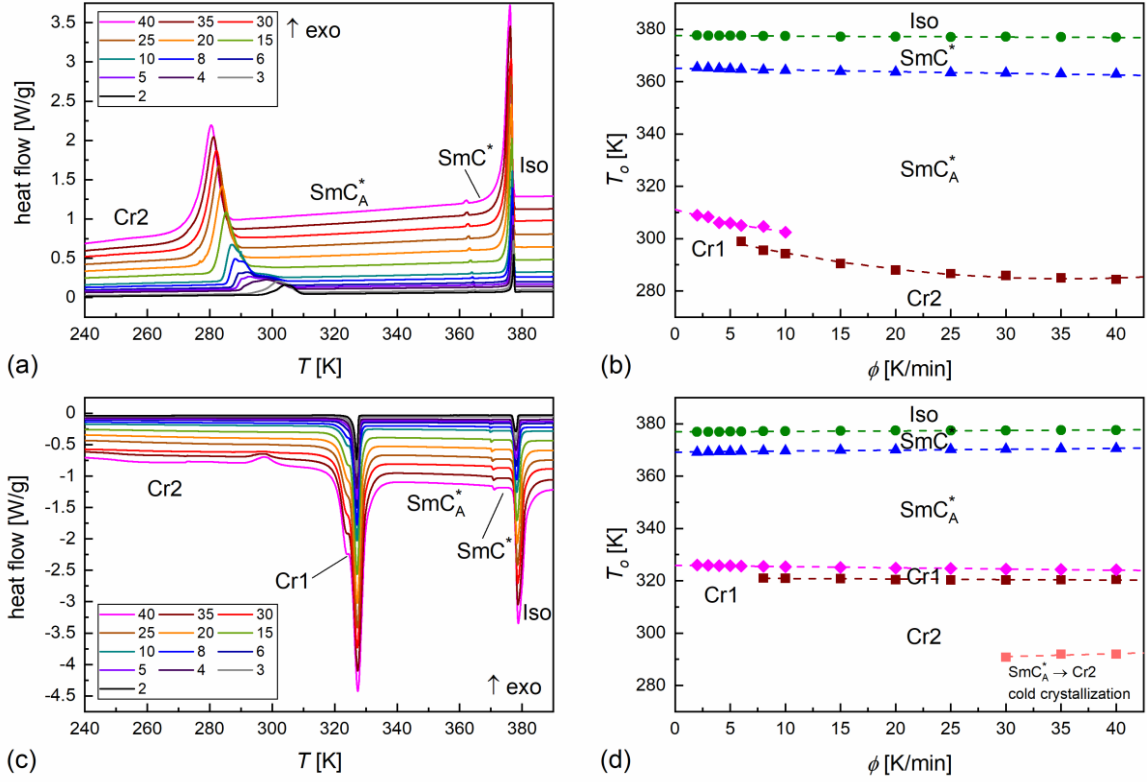

Figure 3. DSC thermograms of 3F2HPhF6 collected on cooling at 2-40 K/min (a), onset transition temperatures as a function of the cooling rate (b), DSC thermograms collected on heating (c), and onset transition temperatures as a function of the heating rate (d).

The phase sequence of MIX2HF6, determined from DSC thermograms (Figure 4a,b) at the limit of zero cooling rate, is Iso (395.9 K) SmA* (386.9 K) SmC* (373.6 K) SmC$_A$* (286.7 K) Cr3. The crystallization is only partial and the step-like anomaly with an inflection at $T_g$ = 241.5 K is observed, indicating the glass transition of the remaining SmC$_A$* phase. The glass softening on heating occurs at 244.9 K, and the cold crystallization follows. A few endothermic anomalies related to the melting of the crystal phases are visible on further heating. The melting of the Cr3 phase is observed for the 15-40 K/min heating rates at 286-288 K and melting of the Cr2 phase is observed for 2-25 K/min at 292-296 K. For all heating rates, the Cr1 phase is observed, which melts at 298-307 K. Differences between the melting temperatures indicate that actually, the mixture of Cr1, Cr2 and Cr3 with various mass ratio is formed, depending on the cooling/heating rate. Additionally, minor anomalies from the melting of the crystal phases denoted as Cr' and Cr'' are visible for the 2, 3 K/min heating rates, with the onset temperatures of 325-326 K and 328-329 K, respectively. The phase sequence in the limit of the zero heating rate is: Cr2 (295.2 K) Cr1 (308.0 K) [Cr' (~325 K) Cr'' (~328 K)] SmC$_A$* (375.6 K) SmC* (386.2 K) SmA* (395.9 K) Iso. The Cr3 → Cr2 transition is not visible in the DSC results. This transition possibly occurs simultaneously with the cold crystallization, or the melt crystallization during slower cooling already includes the formation of Cr2.

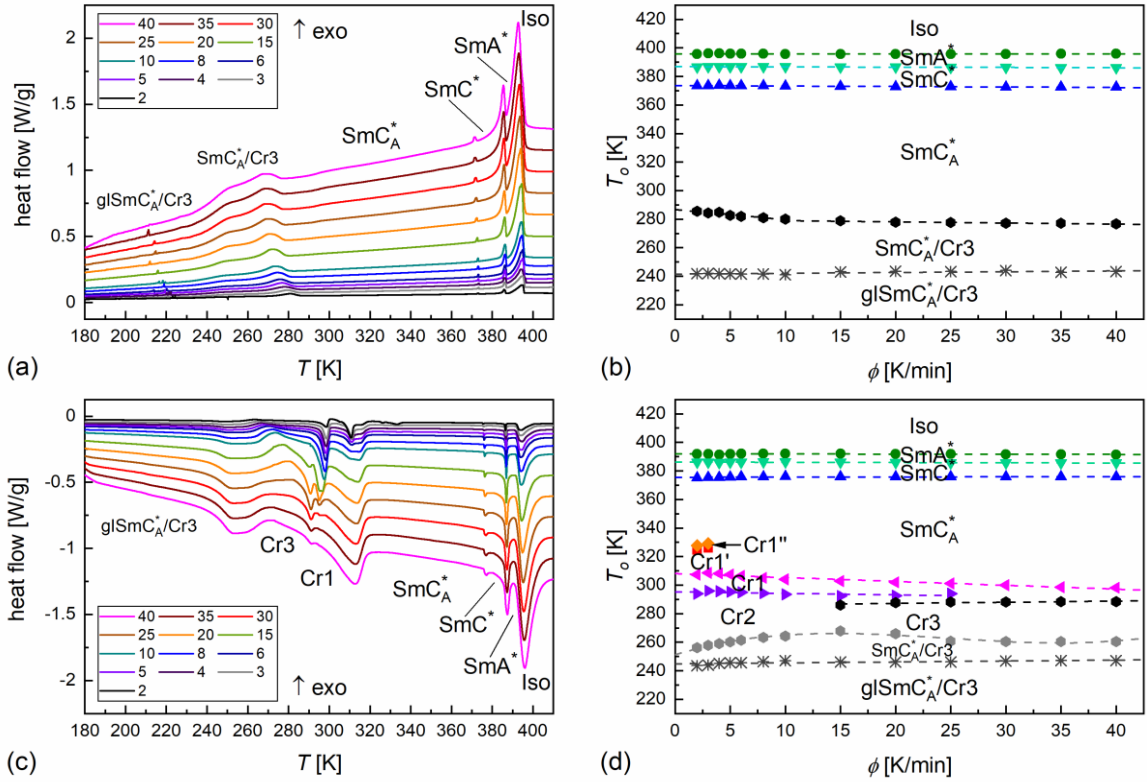

Figure 4. DSC thermograms of MIX2HF6 collected on cooling at 2-40 K/min (a), onset transition temperatures as a function of the cooling rate (b), DSC thermograms collected on heating (c), and onset transition temperatures as a function of the heating rate (d).

The POM textures of MIX2HF6 are partially homeotropic and partially strongly defective. For the textures collected during cooling and heating at 5 K/min (Figure S3), the numerical analysis was performed not only for the whole textures, but also for the homeotropically oriented fragment (indicated for 273 K in Figure S3a), to visualize better the helix inversion and beginning of crystallization. The Cr3 → Cr1 (or Cr2 → Cr1) transition is not visible during heating. The crystallization and glass transition during fast cooling at 30 K/min (Figure S4) show a continuous increase and decrease in the luminance, respectively. The glass softening and cold crystallization during heating do not show clearly in the luminance vs. temperature plot. The Cr3 → Cr1 and Cr1 → SmC$_A$* transitions are visible, which is in accordance with the DSC results for the 30 K/min rate. However, an additional transition occurs at 330 K, attributed to the melting of the Cr1'' phase, observed in the DSC thermograms only for the 2-3 K/min rates. The irregular presence of the Cr1' and Cr1'' phases may be related to the local deviations from the equimolar ratio of components in the solid state, resulting in the small fractions of crystal phases with elevated melting temperatures. The texture in the SmC$_A$* phase after melting of the crystal phase is mainly strongly-defected fan-shaped, without homeotropic fragments. Therefore, the helix inversion is not visible during heating. The polymorphism of the smectic phases is summarized in Table 1. The effect of the sample contained is visible, as the phase transition temperatures determined by POM for a thin sample are shifted to higher temperatures compared to the DSC results. The identification of the smectic phases agrees with the XRD patterns (Section 3.2) and is confirmed by the dielectric spectra (Section 3.3), although for MHPOBC, the BDS method suggests the presence of two other smectic phases besides SmC* in a narrow temperature range between SmA* and SmC$_A$*.

Table 1. Polymorphism of the smectic phases of MIX2HF6 and its components. The presence and absence of a smectic phase are indicated by ● and -, respectively. The first row (normal font) contains the transition temperatures in K obtained by POM (5 K/min), the second row (**bold**) contains the transition temperatures in K obtained by DSC (extrapolated to 0 K/min), the third row (*italics*) contains the enthalpy changes in kJ/mol determined by DSC.

| sample | SmX$_A$* |  | SmC$_A$* |  | SmC* |  | SmA* |  | Iso |
|---|---|---|---|---|---|---|---|---|---|
|  |  |  |  | cooling |  |  |  |  |  |
| MHPOBC | ● | **337.3** *1.5* | 339 ● | **386.6** *>0.1* | 394[a] ● | **388.3** *>0.1* | 394[a] ● | **418.2** *6.0* | 424 ● |
| 3F2HPhF6 | - |  | ● | **365.1** *>0.1* | 366 ● |  |  | - | **377.6** *6.8* | 380 ● |
| MIX2HF6 | - |  | ● | **373.6** *>0.1* | 377 ● | **386.9** *1.1* | 388 ● | **395.9** *4.6* | 394 ● |
|  |  |  |  | heating |  |  |  |  |  |
| MHPOBC | - |  | ● | **385.0** *>0.1* | 394[a] ● | **389.3** *>0.1* | 394[a] ● | **416.6** *5.9* | 424 ● |
| 3F2HPhF6 | - |  | ● | **369.2** *>0.1* | 371 ● |  |  | - | **376.9** *6.8* | 380 ● |
| MIX2HF6 | - |  | ● | **375.6** *>0.1* | 377 ● | **386.2** *0.9* | 388 ● | **395.9** *4.8* | 394 ● |

[a] only the textures collected in 394 K during cooling and heating were attributed to SmC*

### 3.2. Structure of smectic phases

The XRD patterns of the smectic phases (Figure 5a) are characterized by the sharp diffraction peak or peaks located at low 2θ angles. The peak position is related to the smectic layer spacing by the Bragg equation [23]:

$$n\lambda = 2d \sin\theta, \quad (1)$$

where $n$ is the order of the peak ($n = 1, 2, 3…$ – the number of visible peaks of the order $n > 1$ varies between the compounds and smectic phases), $\lambda = 1.540562$ Å is the wavelength of characteristic CuK$\alpha_1$ radiation [38], $d$ is the layer spacing, and $\theta$ is the peak position corrected for the systematic shift (in this study, results for the SRM675 reference material were used for this correction). The XRD patterns at higher 2θ angles depend on the intra-layer order in the smectic phase [23]. For the smectics

A and C, the intra-layer positional order is short-range, described by the average intermolecular distance $w$ and correlation length $\xi$. It corresponds to the wide maximum in the XRD patterns located around $2\theta \approx 18\text{-}20°$, defined by the Lorentz peak function [39]:

$$I(q) \propto \frac{1}{1+\xi^2(q-q_0)^2}, \quad (2)$$

where $q = 4\pi/\sin\theta$ is the scattering vector and $q_0 = 2\pi/w$ is the peak position. Upon transition to the hexatic phase, the diffuse maximum becomes narrower and with a sharper peak [23]. For MIX2HF6, sharpening of the peak at $2\theta \approx 19°$ is observed at 298 K. However, it is interpreted as the beginning of crystallization, because two other sharp peaks at $2\theta \approx 16°$ and $23°$ appear also in this temperature.

The smectic layer spacing in MIX2HF6 is between the results for the pure compounds (Figure 5b). The layer shrinkage at the SmA* → SmC* transition equals 3.4(3)%, and the overall layer shrinkage down to 298 K equals 11.2(2)%. For comparison, the overall layer shrinkage in the smectic phases of MHPOBC and 3F2HPhF6 equals 5.2(3)% and 0.5(2)%, respectively. The wide maximum at higher angles was fitted using Equation (2) with the linear background added. The correlation length in MIX2HF6 reaches almost 6 Å at 303 K before crystallization in 298 K (Figure 5c). The maximal $\xi$ value obtained for 3F2HPhF6 was about 5 Å at 323 K [24], and for MHPOBC, it was ca. 7 Å at 348 K [15]. Thus, the values for the mixture are intermediate between these for pure components. The distance $w$ equals 4.6-4.8 Å, as usually obtained for liquid crystals [24]. The $\xi/w$ ratio does not exceed 2. Therefore, the local positional order in MIX2HF6 includes only the next neighbors.

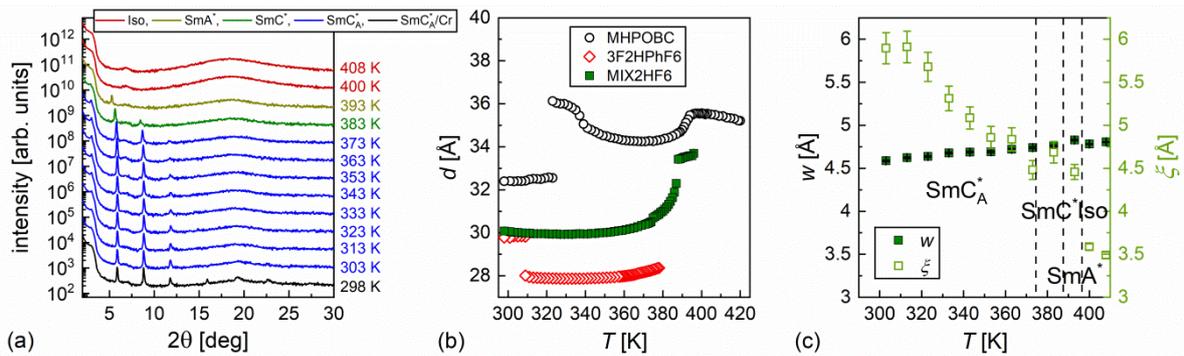

Figure 5. XRD patterns of the MIX2HF6 mixture (a), the smectic layer spacing $d$ in MIX2HF6 and its components (b), and average intermolecular distance $w$ and correlation length $\xi$ of the short-range order in MIX2HF6.

### 3.3. Dielectric relaxation processes

The dielectric spectra show the complex dielectric permittivity $\varepsilon^*$ as a function of the frequency $f$ of the external electric field. The models used for fitting the experimental $\varepsilon^*(f)$ dependences in this study are the Debye [40], Cole-Cole [41] and Havriliak-Negami [42] models, together with the contribution of the ionic conductivity [40]:

$$\varepsilon^*(f) = \varepsilon'(f) - i\varepsilon''(f) = \varepsilon_\infty + \sum_j \frac{\Delta\varepsilon_j}{\left(1+(2\pi if\tau_j)^{1-a_j}\right)^{b_j}} - \frac{iS}{(2\pi f)^{n_S}}, \qquad (3)$$

where $\varepsilon'$ and $\varepsilon''$ are the real and complex component of $\varepsilon^*$ (corresponding to dielectric dispersion and absorption, respectively), $\varepsilon_\infty$ is the $\varepsilon'$ value in the limit of high frequencies, $\Delta\varepsilon$ is the dielectric increment of each process, $\tau$ is the relaxation time of each process, $S$ and $n_S$ are parameters describing the background in $\varepsilon''$ at low frequencies. If $n_S = 1$, then $S = \sigma/\varepsilon_0$, where $\sigma$ is the ionic conductivity and $\varepsilon_0$ is the vacuum permittivity. The parameters $a$ and $b$ describe the shape of the step in $\varepsilon'$ and peak in $\varepsilon''$ originating from each relaxation process. For the Debye model, $a = 0$ and $b = 1$. For the Cole-Cole model, $a \in (0,1)$ and $b = 1$. For the Havriliak-Negami model, $a \in (0,1)$ and $b \in (0,1)$.

The presence of some relaxation processes facilitates the identification of the smectic phases. The soft mode (SM) is related to collective fluctuations of the tilt angle value and is observed in the SmA* and SmC* phases. On cooling, $\tau$ and $\Delta\varepsilon$ of SM increase in SmA* and decrease in SmC*, showing the characteristic V-dependence on temperature [43]. The Goldstone mode (GM) is related to collective fluctuations of the tilt azimuth around the tilt cone and is observed in the SmC* phase. It shows $\Delta\varepsilon$ much larger than SM [43]. In the SmC$_A$* phase, there are two GM-like phasons P$_L$ and P$_H$, related to collective fluctuations of the tilt azimuth which are, respectively, in-phase and anti-phase in the neighbor smectic layers. P$_L$ has higher $\tau$ than P$_H$ and it can overlap with the molecular s-process (rotations around the short molecular axis) [44,45].

The representative fits of Equation (3) to the experimental BDS spectra of MIX2HF6 are shown in Figure 6. The $\Delta\varepsilon$, $\tau$ and $\sigma$ values, determined as a function of temperature, are presented in Figures 7, 8, 9 for MHPOBC, 3F2HPhF6 and MIX2HF6, respectively. The SM mode with $\tau$ and $\Delta\varepsilon$ increasing on cooling confirms the SmA* phase in MHPOBC and MIX2HF6. The GM mode with high $\Delta\varepsilon$ values confirms the SmC* phase and the pair of weaker phasons P$_L$ and P$_H$ confirm the SmC$_A$* phase in MIX2HF6 and both components. GM is observed also in the SmC$_{FI1}$* phase [45], which can be present for MHPOBC. Thus, the strongest value of GM's $\Delta\varepsilon$ at 391 K is attributed to SmC*, while lowered $\Delta\varepsilon$ values at 389-390 K can be attributed to SmC$_{FI1}$*. From the high-temperature side, the relatively strong SM may indicate the SmC$_\alpha$* phase [45,46]. The final phase sequence in MHPOBC would be SmA*, SmC$_\alpha$*, SmC*, SmC$_{FI1}$*, and SmC$_A$*. On the other hand, the DSC thermograms show only one phase between SmA* and SmC$_A$*. We refer to this phase as SmC*, because this phase is the most probable based on the BDS results. The SmC$_A$* → SmX$_A$* transition in the supercooled MHPOBC results in an increase and decrease of $\Delta\varepsilon$ of P$_L$ and P$_H$, respectively (Figure 7a), and in the increased slope of the activation plot of $\tau$ of both phasons (Figure 7b). For some compounds, the additional $P_{hex}$ process arises at the low-frequency side of $P_H$ in the $SmX_A^*$ phase [28,47]. However, it was not visible for MHPOBC. Compared to MIX2HF6 and 3F2HPhF6, which show comparable relaxation times, the GM process is much slower and the $P_L$ process is much faster for MHPOBC. The $P_L$ process, probably due to overlapping with the molecular s-process, shows the linear dependence in the activation plot. The activation energy $E_a$, determined by fitting the Arrhenius formula, is ca. 110 kJ/mol for all samples. For the $P_H$ process, the difference in the relaxation times between the mixture and its components is insignificant. The ionic conductivity is the highest for MIX2HF6, slightly smaller for 3F2HPhF6 and the lowest for MHPOBC.

The MIX2HF6 can be supercooled enough to investigate the next relaxation process on the high-frequency side of the $P_H$ process (Figure 6). It can be interpreted as the molecular l-process (rotations around the long molecular axis) [48]. However, for the glassforming materials, this process becomes the collective α-process on approaching the glass transition temperature $T_g$ [40]. The α-relaxation time usually shows the nonlinear dependence in the activation plot, described by the Vogel-Fulcher-Tammann formula [40,49]:

$$\tau_\alpha(T) = \tau_\infty \exp\left(\frac{B}{T-T_V}\right), \tag{4}$$

where $\tau_\infty$ is a pre-exponential constant, $T_V$ is the Vogel temperature and $B$ is a parameter with an unit of temperature. For the Arrhenius dependence, $T_V = 0$ and $B = E_a/R$. The α-relaxation in MIX2HF6 cannot be investigated in the vicinity of $T_g$ due to crystallization and overlapping with the cr-process from the crystal phase. $T_g$ is defined as the temperature where $\tau_\alpha = 100$ s [49]. The α-relaxation time determined in higher temperatures, down to 259 K, shows the linear dependence (Figure 9). Only two points collected upon heating after fast cooling (Figure 9h) at 257 K and 255 K deviate from the Arrhenius formula, but at the same time they do not follow the VFT equation. Thus, the α-relaxation time cannot be used for determination of the fragility index $m_f$, defined as [49]:

$$m_f = \left.\frac{d\log_{10}\tau_\alpha}{d(T_g/T)}\right|_{T=T_g} \tag{5}$$

and describing deviation of $\tau_\alpha(T)$ from the Arrhenius formula. For the strong glassformers, $\tau_\alpha(T)$ follows the Arrhenius formula and $m_f = 16$. For very fragile glassformers, which $\tau_\alpha(T)$ dependence shows strongly nonlinear dependence in the activation plot, $m_f$ may reach more than 200. The Arrhenius behavior of $\tau_\alpha$ could be interpreted as MIX2HF6 being a strong glassformer, but the deviations from the Arrhenius formula may appear in lower temperatures (as suggested by results from 257 K and 255 K), where $\tau_\alpha$ was not investigated.

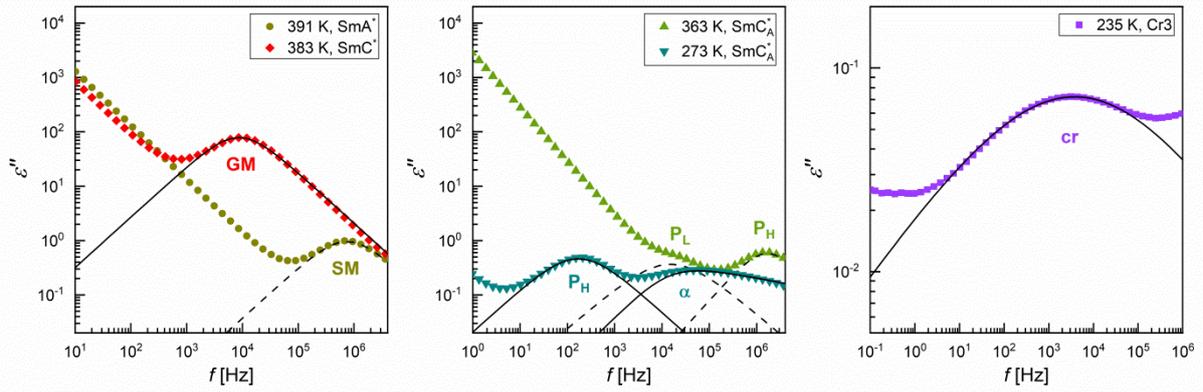

Figure 6. Absorption parts of selected BDS spectra of MIX2HF6 collected on slow cooling (points) and fitting results of Equation (3) (lines). The conductivity part was omitted.

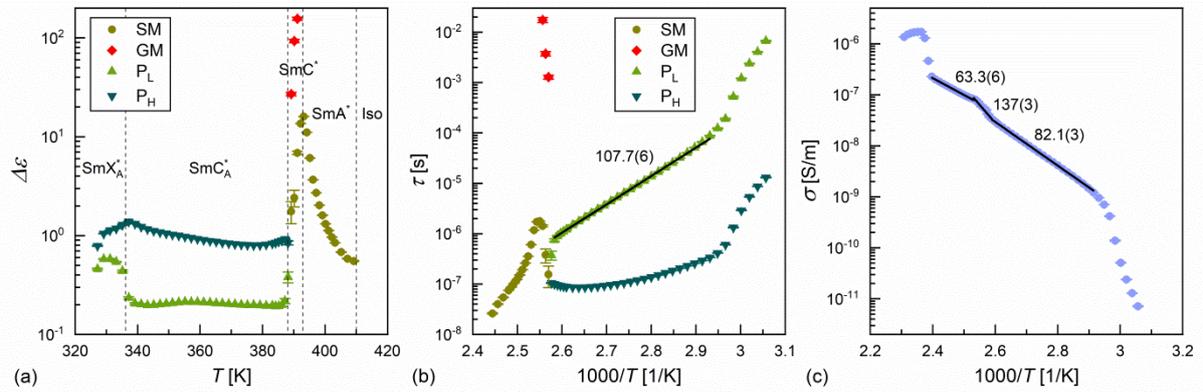

Figure 7. BDS results for MHPOBC obtained on slow cooling: dielectric increment (a), activation plot of relaxation times (b), and activation plot for ionic conductivity (c). The activation energies in (b) and (c) are in kJ/mol.

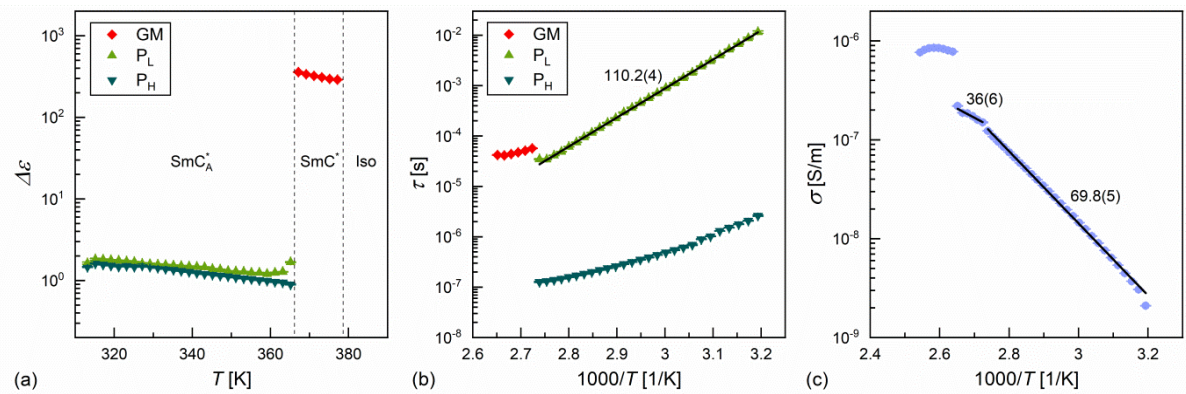

Figure 8. BDS results for 3F2HPhF6 obtained on slow cooling: dielectric increment (a), activation plot of relaxation times (b), and activation plot for ionic conductivity (c). The activation energies in (b) and (c) are in kJ/mol.

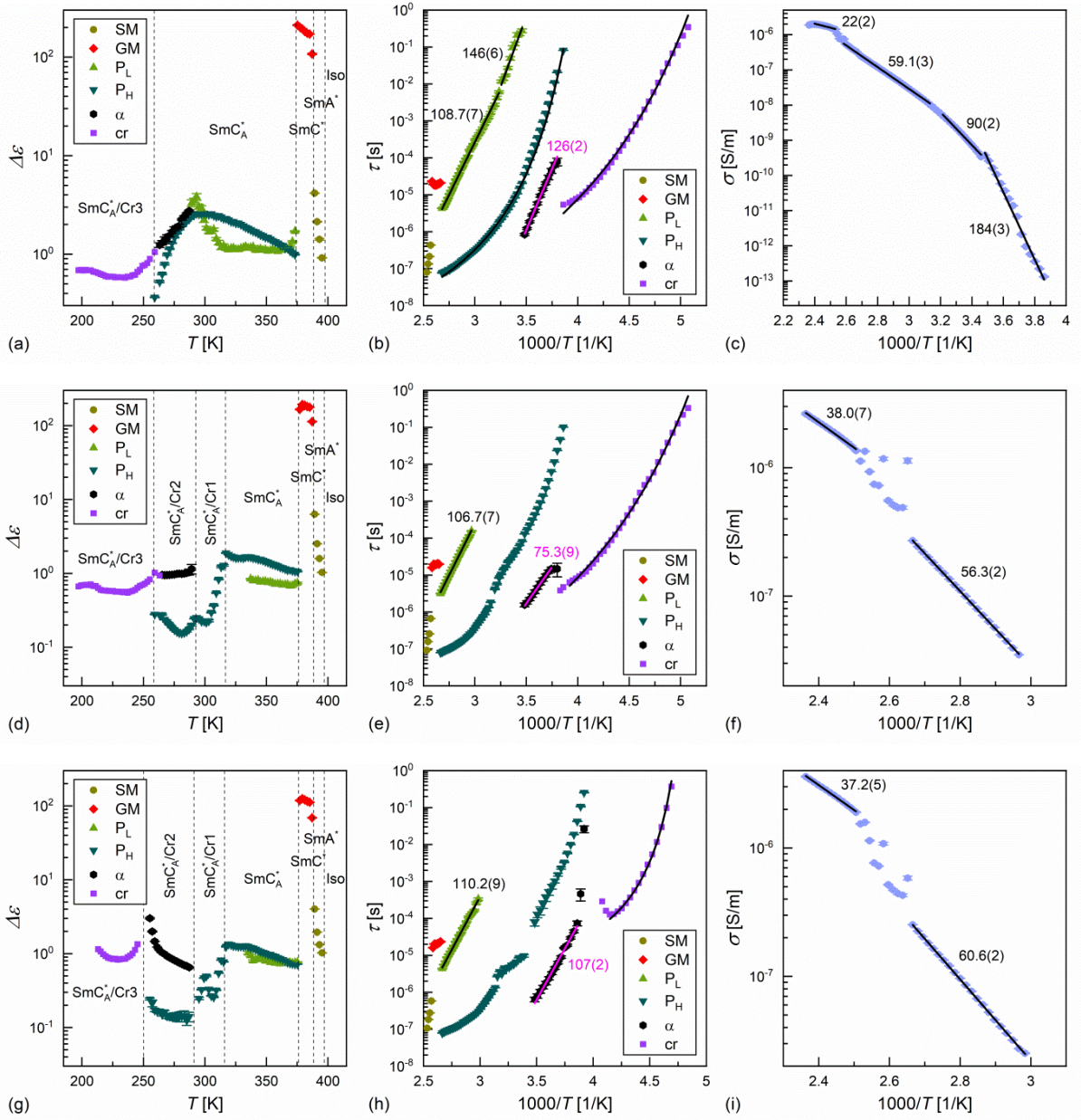

Figure 9. BDS results for MIX2HF6 obtained on slow cooling (a-c), heating after slow cooling (d-f), and heating after fast cooling at 10 K/min (g-i): dielectric increment (a,d,g), activation plot of relaxation times (b,e,h), and activation plot for ionic conductivity (c,f,i). The activation energies in (b,c,e,f,h,i) are in kJ/mol.

There is an alternative way of determination of $m_f$, based on the DSC thermograms [50]:

$$m_f = \frac{cT_g \Delta C_p}{\Delta H_m}, \qquad (6)$$

where $\Delta C_p$ is the jump in the heat capacity at $T_g$ and $\Delta H_m$ is the enthalpy of melting. Based on the experimental data for more than 40 glassformers, the empirical $c$ parameter was usually ca. 56 [50]. The modulated-temperature DSC results were used to determine the $m_f$ parameter of MIX2HF6 (Figure 10). The MTDSC technique separates the non-reversing and reversing parts of the heat flow [51]. For MIX2HF6, these parts of the heat flow correspond to the cold crystallization and glass transition, overlapping in the standard DSC thermogram. The glass transition occurs at $T_g$ = 245.3 K, with $\Delta C_p$ = 0.16 kJ/(mol·K). As the melting enthalpy, the summed melting enthalpies of the Cr1 and Cr2 phases were included, giving $\Delta H_m$ = 9.2 kJ/mol. Using Equation (6) and $c$ = 56, one obtains $m_f$ = 233, which is an exceptionally high value [49]. However, for a few substances presented in [50], $c$ deviated strongly from 56, which can also be the case for MIX2HF6. Recently, we published the DSC and BDS results for the 3F7FPhH6 compound [29]. 3F7FPhH6 has a molecular structure similar to MIX2HF6 components and it forms the SmC$_A$* glass. The fragility index obtained from the α-relaxation time is $m_f$ = 72. Using the $T_g$ = 233.9 K, $\Delta C_p$ = 0.18 kJ/(mol·K), and $\Delta H_m$ = 15.7 kJ/mol values determined from the DSC thermograms of 3F7FPhH6, one can conclude that the kinetic fragility agrees with the thermodynamic fragility for $c$ = 26. If the same $c$ parameter is applied for MIX2HF6, then $m_f$ = 108, a more reasonable value, within the 72-149 range obtained for similar compounds [28-30,48].

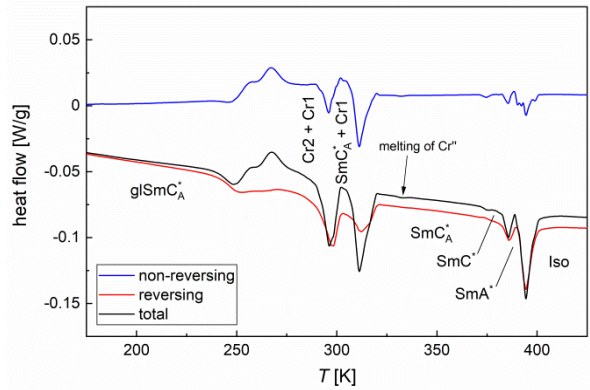

Figure 10. MTDSC thermograms collected for MIX2HF6 during heating with the 3 K/min rate after quenching to 153 K.

Coming back to the BDS spectra of MIX2HF6, one can see that the P$_H$-relaxation time has a nonlinear dependence in the activation plot and it is expected to reach 100 s at a temperature only slightly higher than $T_g$. Moreover, P$_H$ appears at much lower frequencies than the cr-process from the crystal phase, therefore on cooling it is less affected by crystallization than the α-process. The P$_H$-relaxation time well fits the VFT formula (Figure 9b). The approximate $T_g$ obtained from the fitting parameters is 249 K, close to the calorimetric 245.3 K. The fragility index obtained from the P$_H$-relaxation time is 100, also surprisingly close to calorimetric 108. However, it does not mean that P$_H$ can always be used to determine $m_f$. In publication [48], where both P$_H$ and α-relaxation times were analyzed by the VFT formula, the obtained $T_g$ values differed only by 2 K. At the same time, $m_f$ were less consistent (117 and 149, respectively).

The last relaxation process in MIX2HF6 is the cr-process from the Cr3 phase, absent in Cr2 and Cr1. Its relaxation time also follows the VFT equation. While the fitting parameters are similar for slow cooling and subsequent heating (Figure 9b,e), they are different for heating after fast cooling (Figure 9h) – one can see that in the latter, the cr-relaxation time is longer and diverges at higher temperatures. The relaxation process in the crystal phase can be interpreted as the conformational changes, corresponding to the conformationally-disordered crystal phase (CONDIS phase), which can undergo the glass transition [52-54]. Thus, the Cr3 phase is probably the CONDIS phase, while in Cr2 and Cr1 the conformational disorder is lower or absent. If $T_g$ of Cr3 is defined as the temperature where the cr-relaxation time reaches 100 s, then the VFT fits give $T_g$ = 188 K, $m_f$ = 52-54 for slow cooling and heating after slow cooling, and $T_g$ = 209 K, $m_f$ = 161 for heating after fast cooling. The differences in the cr-process in the slowly and fast-cooled sample may result from a larger fraction of the remaining SmC$_A$* phase during fast cooling and the formation of both the Cr3 and Cr2 phases during slow cooling. Both can influence the microstructure of the Cr3 phase and, consequently, affect the cr-relaxation in Cr3 due to the confinement effect [55,56].

### 3.3. Non-isothermal crystallization

The melt crystallization of MIX2HF6 and its components was investigated by DSC for the cooling rates in the 2-40 K/min range (Figures 2-4). The relative crystallization degree $x$ was determined as an integral of the heat flow $\Phi$ between temperatures $T_{start}$ and $T_{end}$, indicating the temperature range of the exothermic anomaly related to crystallization [57]:

$$x(T) = \frac{\int_{T_{start}}^{T} \Phi(T)dT}{\int_{T_{start}}^{T_{end}} \Phi(T)dT}. \qquad (7)$$

The data were analyzed with the isoconversional method (Figures 11-13), which enables the determination of the effective activation energy $E_{eff}$ as a function of $x$ and $T$ [57-60]. This method is based on the formula [58,59]:

$$\frac{dx(t)}{dt} = f(x)A \exp\left(-\frac{E_{eff}}{RT}\right), \qquad (8)$$

where $f(x)$ is the transition model and $A$ is a pre-exponential factor. The logarithm of the crystallization rate $dx/dt$ is plotted against inverted temperature $T_x$ for a selected $x$, for all cooling rates. The slope of such an activation plot is equal to $-E_{eff}/R$. The obtained $E_{eff}$ can be plotted against temperature if the average temperature for each range of the linear fit is used [58].

The isoconversional analysis for MHPOBC reveals three crystallization regimes (Figure 11). The effective activation energies $E_{eff1}$ and $E_{eff2}$, which correspond mainly to the melt crystallization above the room temperature, are negative and equal to –(382-329) kJ/mol and –(171-84) kJ/mol, respectively. The negative values mean that the crystallization kinetics depends mainly on nucleation, i.e., the thermodynamic driving force controls it. Below the room temperature, the effective activation energy $E_{eff3}$ takes values between −121 kJ/mol and 37 kJ/mol. Positive $E_{eff3}$ values, obtained for the intermediate degree of crystallization, mean that in lower temperatures, the diffusion rate has a bigger impact on the crystallization kinetics than above the room temperature, where the impact of the nucleation rate was the most significant.

The results for 3F2HPhF6 (Figure 12) show four regimes of crystallization. Above the room temperature, $E_{eff1}$ = −(66-11) kJ/mol, around 290 K, $E_{eff2}$ = −(170-134) kJ/mol, and below 285 K, $E_{eff3}$ changes between −81 kJ/mol and 22 kJ/mol. Thus, the crystallization kinetics above the room temperature is controlled by nucleation; the influence of the nucleation rate increases down to 290 K, and in lower temperatures, the diffusion rate has a gradually growing impact. By comparing these results with the phase sequence of 3F2HPhF6 for various cooling rates (Figure 3), one can conclude that $E_{eff1}$ and $E_{eff3}$ correspond to the SmC$_A$* → Cr1 and SmC$_A$* → Cr2 transitions, respectively. At the same time, $E_{eff2}$ describes the temperature range where the mix of Cr1 and Cr2 is formed. For the final stages of crystallization at 30-40 K/min, one can observe the fourth regime, where the activation energy is negative, $E_{eff4}$ = −(80-65) kJ/mol (Figure 12b).

For MIX2HF6, the isoconversional analysis indicates two regimes, both described by the negative effective activation energies: $E_{eff1}$ = −(122-92) kJ/mol for the 2-10 K/min cooling rates and $E_{eff2}$ = −(165-119) kJ/mol for the 15-40 K/min cooling rates (Figure 13). In the beginning stages of crystallization ($x$ = 0.1-0.2), the activation energy $E_{eff2}$ corresponds to the crystallization kinetics for all applied cooling rates. $E_{eff1}$ and $E_{eff2}$ cover approximately the same temperature range, except the region below 265 K, where only $E_{eff2}$ describes the crystallization kinetics. The negative $E_{eff1}$ and $E_{eff2}$ values for MIX2HF6 show that its melt crystallization is controlled mainly by nucleation, while the diffusion rate does not have a significant impact, unlike the pure compounds, where $E_{eff}$ was also generally negative, but positive $E_{eff}$ values were obtained for certain temperatures.

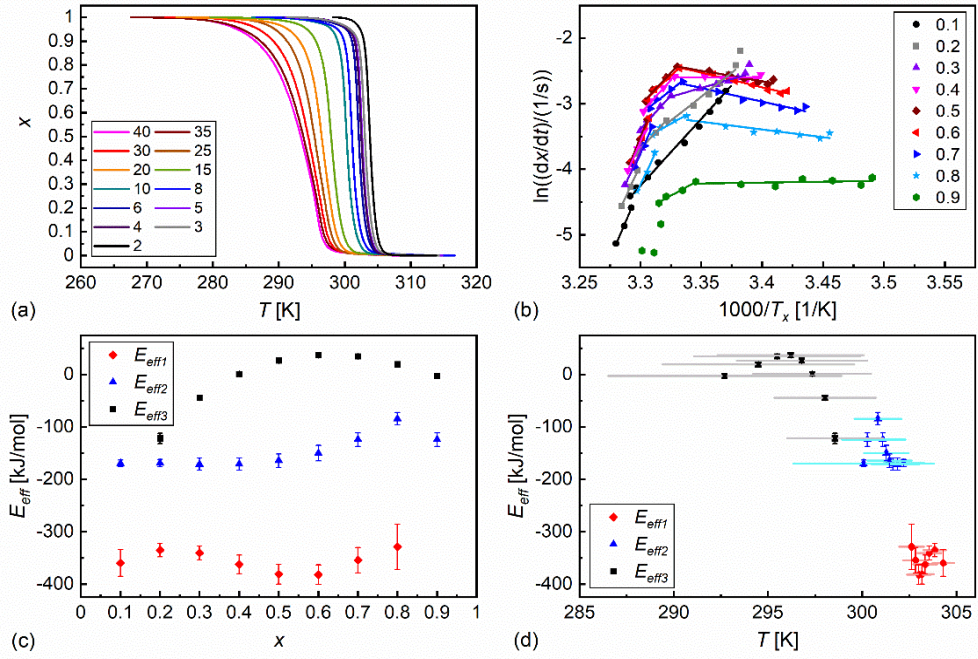

Figure 11. Crystallization degree vs. temperature of MHPOBC (a), activation plot based on isoconversional method (b), effective activation energy vs. crystallization degree (c) and temperature (d).

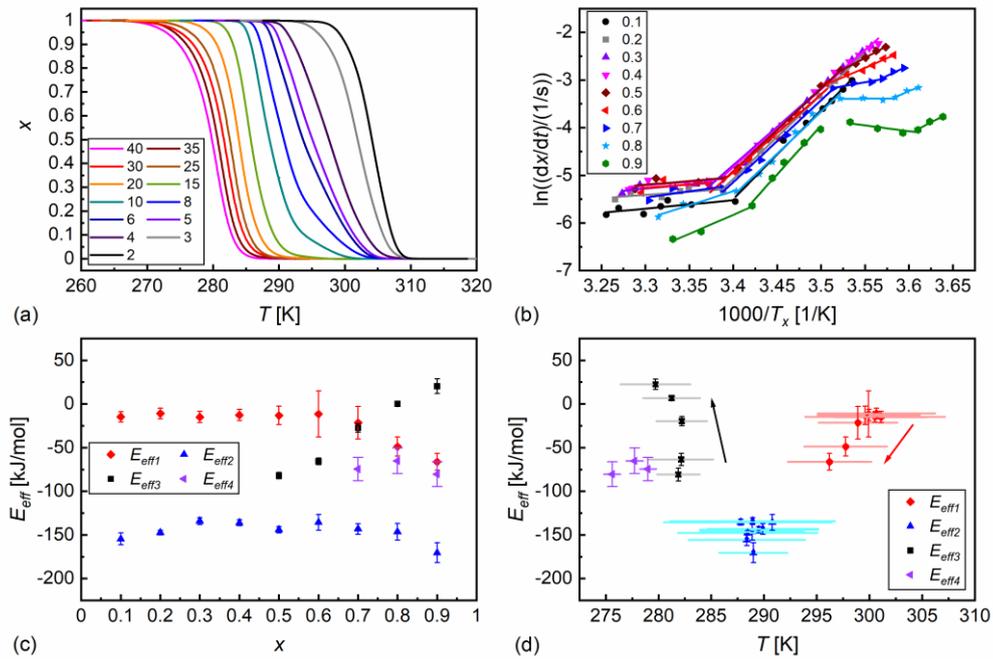

Figure 12. Crystallization degree vs. temperature of 3F2HPhF6 (a), activation plot based on isoconversional method (b), effective activation energy vs. crystallization degree (c) and temperature (d).

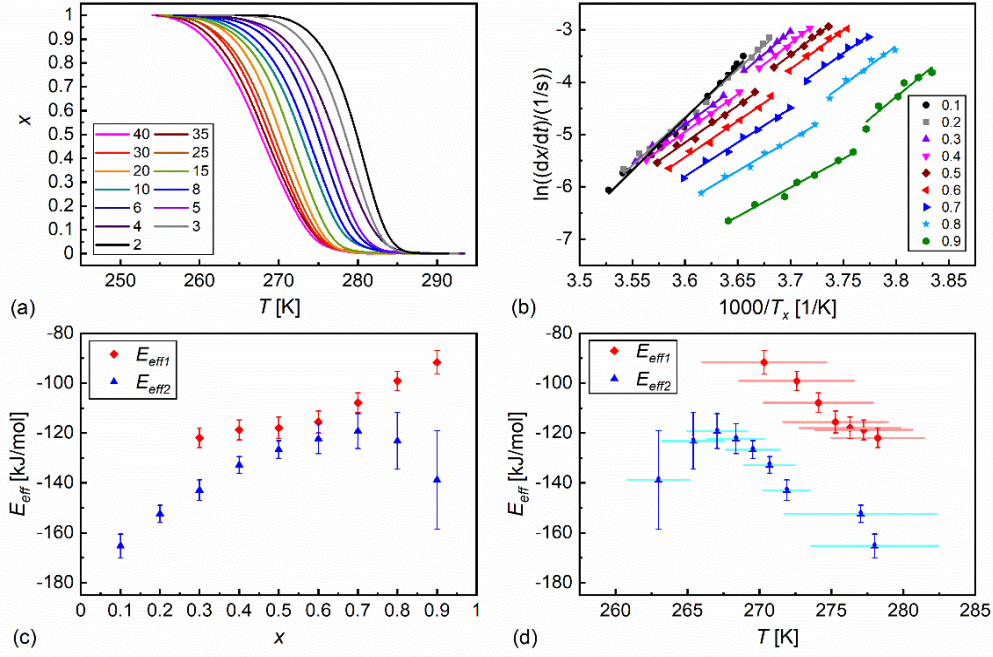

Figure 13. Crystallization degree vs. temperature of MIX2HF6 (a), activation plot based on isoconversional method (b), effective activation energy vs. crystallization degree (c) and temperature (d).

## 3.4. Isothermal crystallization

The isothermal crystallization was investigated by the DSC method. Each sample was heated to the isotropic liquid and cooled at 20 K/min to the crystallization temperature $T_{cr}$. The ranges of $T_{cr}$ were selected based on the DSC results from Figures 2-4. The isothermal DSC thermograms as a function of time are presented in Figure 14a-c. The crystallization degree vs. time was obtained using Equation (7), but the integration was performed over time, not over temperature. The $x(t)$ dependences were fitted by the Avrami model [61,62]:

$$x(t) = 1 - \exp\left(\left(\frac{t-t_0}{\tau_{cr}}\right)^n\right), \qquad (9)$$

where $t_0$ is the initialization time, $\tau_{cr}$ is the characteristic crystallization time and $n$ depends on the type of nucleation (constant number of nuclei or constant nucleation rate) and dimensionality of the crystal growth (shape of crystallites). For the constant number of nuclei, $n$ = 1, 2, 3 indicates the 1D, 2D, and 3D crystal growth. For the constant nucleation rate, the respective $n$ values are 2, 3, 4 [61-63]. Higher $n$ can be interpreted as the sheaf-like growth of crystallites [63]. The results for 3F2HPhF6 (Figure 14d) can be fitted with Equation (9). For MHPOBC, two overlapping crystallization processes are observed (Figure 14e), which requires the fitting of two Avrami models, where $A$ is the fraction of the crystal developed in the 1st stage of crystallization [64]:

$$x(t) = \begin{cases} A\left[1 - \exp\left(-\left(\frac{t-t_{01}}{\tau_1}\right)^{n_1}\right)\right] & \text{for } t_{01} \leq t \leq t_{02} \\ A\left[1 - \exp\left(-\left(\frac{t-t_{01}}{\tau_1}\right)^{n_1}\right)\right] + (1-A)\left[1 - \exp\left(-\left(\frac{t-t_{02}}{\tau_2}\right)^{n_2}\right)\right] & \text{for } t \geq t_{02} \end{cases}. \quad (10)$$

The crystallization of MIX2HF6 occurs in two stages, but the second stage starts after the completion of the first stage (Figure 14f). Therefore, it is possible to fit also their $x(t)$ dependences separately with Equation (9).

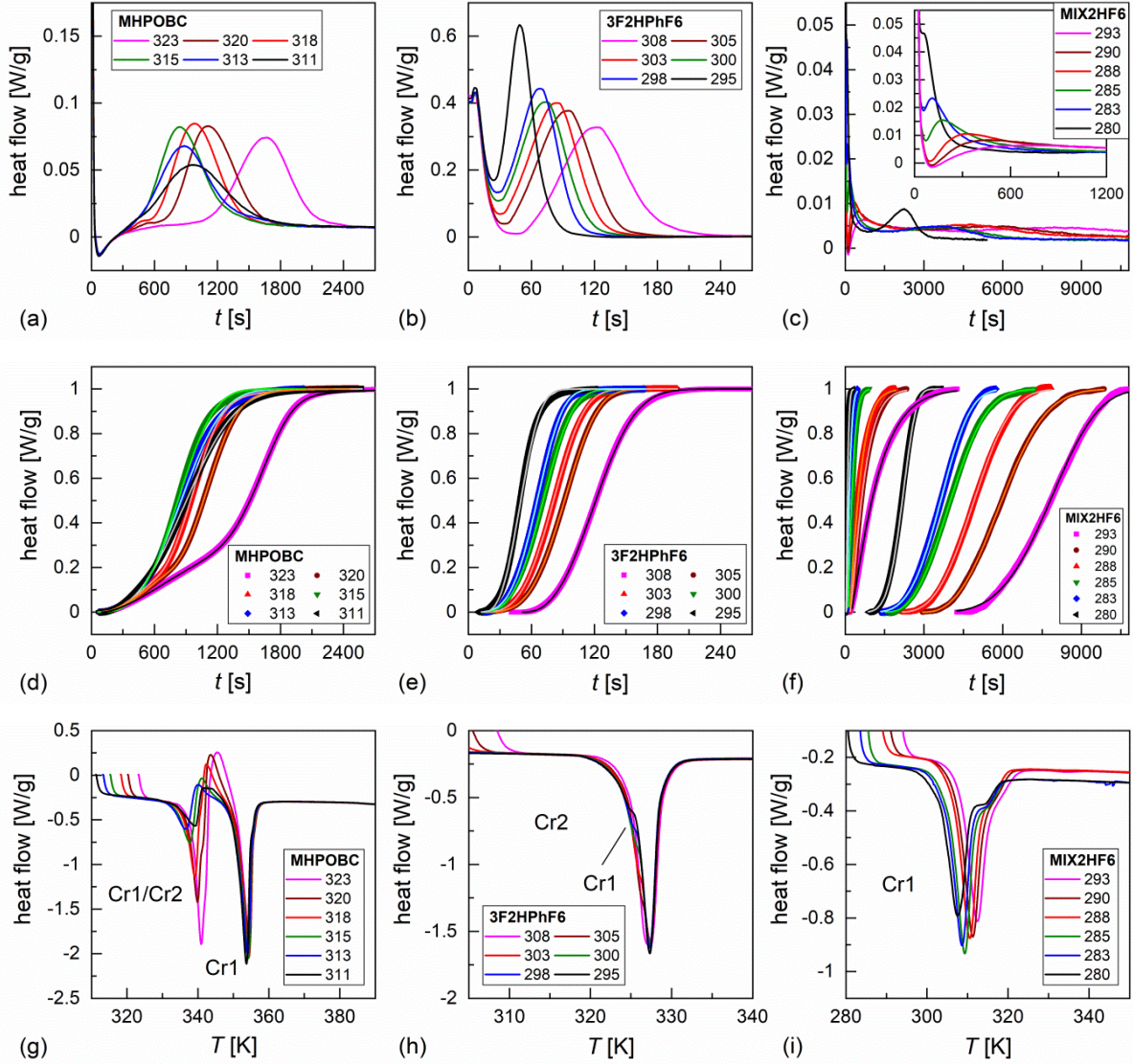

Figure 14. DSC curves collected at selected crystallization temperatures $T_{cr}$ (a-c), isothermal crystallization degree vs. time (d-f), and DSC thermograms collected during heating at 10 K/min after crystallization (g-i). The corresponding $T_{cr}$ values are given in legends in K. The results for MHPOBC, 3F2HPhF6, and MIX2HF6 are shown in panels (a,d,g), (b,e,h), and (c,f,i), respectively.

The isothermal crystallization of MHPOBC results in a mixture of Cr2 and Cr1 phases, with the fraction of the Cr2 phase increasing with increasing $T_{cr}$ (Figure 14g). The 3F2HPhF6 sample consists mostly of the Cr2 phase, which transforms into the Cr1 phase shortly before melting. Only for $T_{cr} = 308$ K, the Cr2 → Cr1 transition is not visible at the slope of the melting anomaly, therefore, the isothermal crystallization at this temperature results in the Cr1 phase (Figure 14h). The structural differences between Cr2 and Cr1 are presumed small, as the Cr2 → Cr1 transition shows a small anomaly in the DSC thermograms. The melting temperature of MIX2HF6 is 303-308 K and increases with increasing $T_{cr}$ (Figure 14i). It is comparable with the melting temperature of the Cr1 phase forming upon heating from the Cr2 or Cr3 phase (Figure 4d). After the isothermal crystallization of MIX2HF6, two closely overlapping melting anomalies are observed, which indicates two crystal phases with slightly different structures or fractions of the components.

The parameters of the Avrami model obtained for MIX2HF6 and its components are presented in Figure 15. The initialization time is shown in the time-temperature-transition (TTT) diagram [TTT] (Figure 15a). In the idealized TTT diagram, $t_0$ is expected to have a minimum at a certain $T_{cr}$. Above and below this temperature, the crystallization kinetics are controlled by the nucleation and diffusion rates. Therefore, $t_0$ decreases and increases with decreasing $T_{cr}$, respectively. For MHPOBC, the minimum of $t_{01}$ is observed at 315 K and of $t_{02}$ at 312 K. For 3F2HPhF6 the shortest $t_0$ is obtained at 298 K. However, the $t_0$ value at 295 K is likely overestimated. For MIX2HF6, only the upper branch of the TTT diagram is observed in the investigated temperature range.

The activation plot of the characteristic crystallization time (Figure 15b) does not show a linear dependence for MHPOBC and 3F2HPhF6. The $\tau_1$ and $\tau_2$ of MHPOBC initially decrease between 323 K and 320 K ($E_{eff} < 0$), then the activation plot is almost horizontal ($E_{eff} \approx 0$). For 3F2HPhF6, $\tau_{cr}$ shows an approximately decreasing trend with decreasing $T_{cr}$ ($E_{eff} < 0$). The $\tau_1$ values for MIX2HF6 show a linear dependence in the activation plot, with $E_{eff} = -141(3)$ kJ/mol. Thus, the first stage of MIX2HF6 crystallization is controlled by the nucleation rate to a larger extent than the crystallization of the pure compounds, and by the second stage of MIX2HF6 crystallization, where $\tau_2$ shows a roughly linear dependence $E_{eff} = -33(6)$ kJ/mol. Comparing the $t_0$ and $\tau_{cr}$ values for all samples, one can say that 3F2HPhF6 shows the fastest crystallization. The 1[st] stage of MIX2HF6 crystallization occurs faster than that of MHPOBC in most cases, but the 2[nd] stage of MIX2HF6 crystallization is much slower than the overall crystallization of MHPOBC. Additionally, the crystallization range of MIX2HF6 is shifted to lower temperatures compared to pure compounds.

The last parameter is the Avrami exponent (Figure 15c). The lowest $n$ values of 1-1.5 are obtained for the 1[st] stage of MIX2HF6 crystallization and are interpreted as mainly 1D crystal growth. For the 1[st] stage of MHPOBC crystallization, $n = 1.5$-2.5, which means a mix of 1D and 2D crystal growth. The 2[nd] stages of MHPOBC and MIX2HF6 crystallization, as well as 3F2HPhF6 crystallization, are characterized by $n = 2.5$-4, which indicates a mix of 2D and 3D growth ($n = 2.5$-3) or mainly 3D growth ($n$ = 3-4) of crystallites. The maximal $n \approx 5$ for the 2nd stage of MIX2HF6 crystallization in 280 K indicates the contribution of the sheaf-like crystal growth.

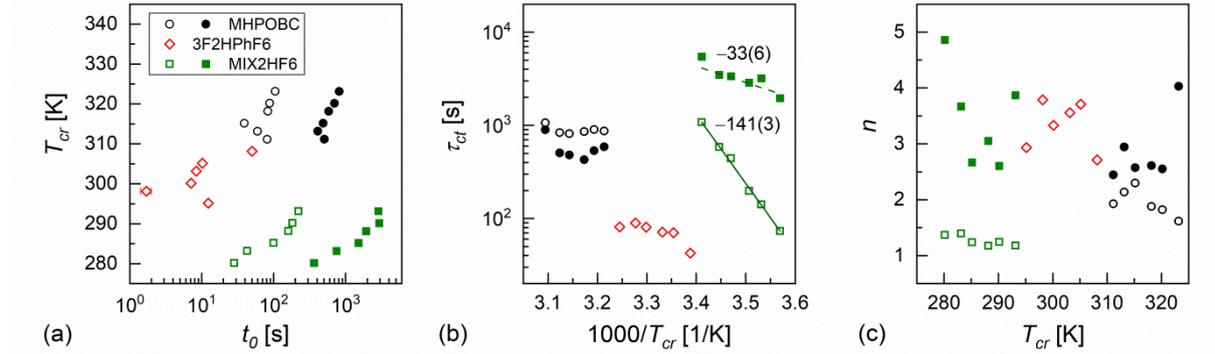

Figure 15. TTT diagram of the initialization time (a), activation plot of the characteristic crystallization time (b), and Avrami parameter as a function of crystallization temperature (c), determined by DSC for MIX2HF6 and its components. The legend in (a) applies to all panels. Open and solid symbols correspond to the first and second crystallization stages, respectively.

## 4. Summary and conclusions

The liquid crystalline compounds with acronyms MHPOBC and 3F2HPhF6 were used to formulate the equimolar mixture, denoted as MIX2HF6. The phase sequences of the pure compounds and their mixture were investigated using the DSC, POM, XRD, and BDS methods. The kinetics of the melt crystallization in non-isothermal conditions (various cooling rates) and isothermal conditions (various crystallization temperatures) was investigated by DSC. The main conclusions regarding the behavior of the mixture and its components are:

- According to the DSC and XRD results, the investigated MHPOBC sample exhibits the SmA*, SmC$_A$*, and hexatic SmX$_A$* phases. At least one smectic phase between SmA* and SmC$_A$* is visible in a narrow temperature range in the DSC thermograms. Based on the BDS spectra, the phase sequence is probably SmA* → SmC$_\alpha$* → SmC* → SmCγ* → SmC$_A$* → SmX$_A$*.
- For 3F2HPhF6, the SmC* and SmC$_A$* phases are observed, which agrees with our previous results.
- The equimolar mixture MIX2HF6 shows the SmA*, SmC*, and SmC$_A$* phases. The experimental results do not indicate the presence of SmC$_\alpha$*, SmCγ* and SmX$_A$*.
- While the pure components crystallize easily during cooling, MIX2HF6 shows only partial crystallization, and the remaining supercooled SmC$_A$* phase forms the glass at $T_g$ = 241.5 K. The glass softening during heating occurs at a slightly higher temperature of ca. 245 K.
- The melt crystallization kinetics is usually controlled by the nucleation rate for the pure components (effective activation energy $E_{eff} < 0$), although in certain conditions the diffusion rate has a dominant contribution to the overall crystallization kinetics ($E_{eff} > 0$). Meanwhile, the melt crystallization of MIX2HF6 is controlled by the nucleation rate in all tested conditions. MIX2HF6 crystallizes at lower temperatures and a lower rate than the pure compounds.

- The melting temperature of MIX2HF6, $T_m$ = 308 K, is also lowered compared to 352 K for MHPOBC and 326 K for 3F2HPhF6. However, in some DSC thermograms of MIX2HF6, small fractions of the crystal phases with higher $T_m$ = 325 K and 328 K are observed. It is attributed to minor inhomogeneities in the mixture's composition, which occur supposedly in the solid state.

The next planned step is to mix the MHPOBC compound with higher 3FmHPhF6 homologs and to investigate of the influence of the $C_mH_{2m}$ chain length on the phase sequence and glassforming properties of the obtained mixtures.

**Acknowledgement:** Aleksandra Deptuch acknowledges the National Science Centre, Poland (grant MINIATURA 7 no. 2023/07/X/ST3/00182) for financial support.

**Conflicts of interest statement:** There are no conflicts to declare.

**Authors' contributions:**

A. Deptuch – conceptualization, investigation, formal analysis, funding acquisition, writing – original draft

A. Paliga – investigation, formal analysis, writing – review and editing

A. Drzewicz – investigation, writing – review and editing

M. Piwowarczyk – resources, writing – review and editing

M. Urbańska – resources, writing – review and editing

E. Juszyńska-Gałązka – investigation, writing – review and editing

# Crystallization kinetics of equimolar liquid crystalline mixture and its components

Aleksandra Deptuch[1,*], Anna Paliga[2], Anna Drzewicz[1], Marcin Piwowarczyk[1], Magdalena Urbańska[3], Ewa Juszyńska-Gałązka[1,4]

[1] Institute of Nuclear Physics Polish Academy of Sciences, Radzikowskiego 152, PL-31342 Kraków, Poland

[2] Faculty of Physics and Applied Computer Science, AGH University of Kraków, Reymonta 19, PL-30059 Kraków, Poland

[3] Institute of Chemistry, Military University of Technology, Kaliskiego 2, PL-00908 Warsaw, Poland

[4] Research Center for Thermal and Entropic Science, Graduate School of Science, Osaka University, 560-0043 Osaka, Japan

*corresponding author, aleksandra.deptuch@ifj.edu.pl

# Supplementary Materials

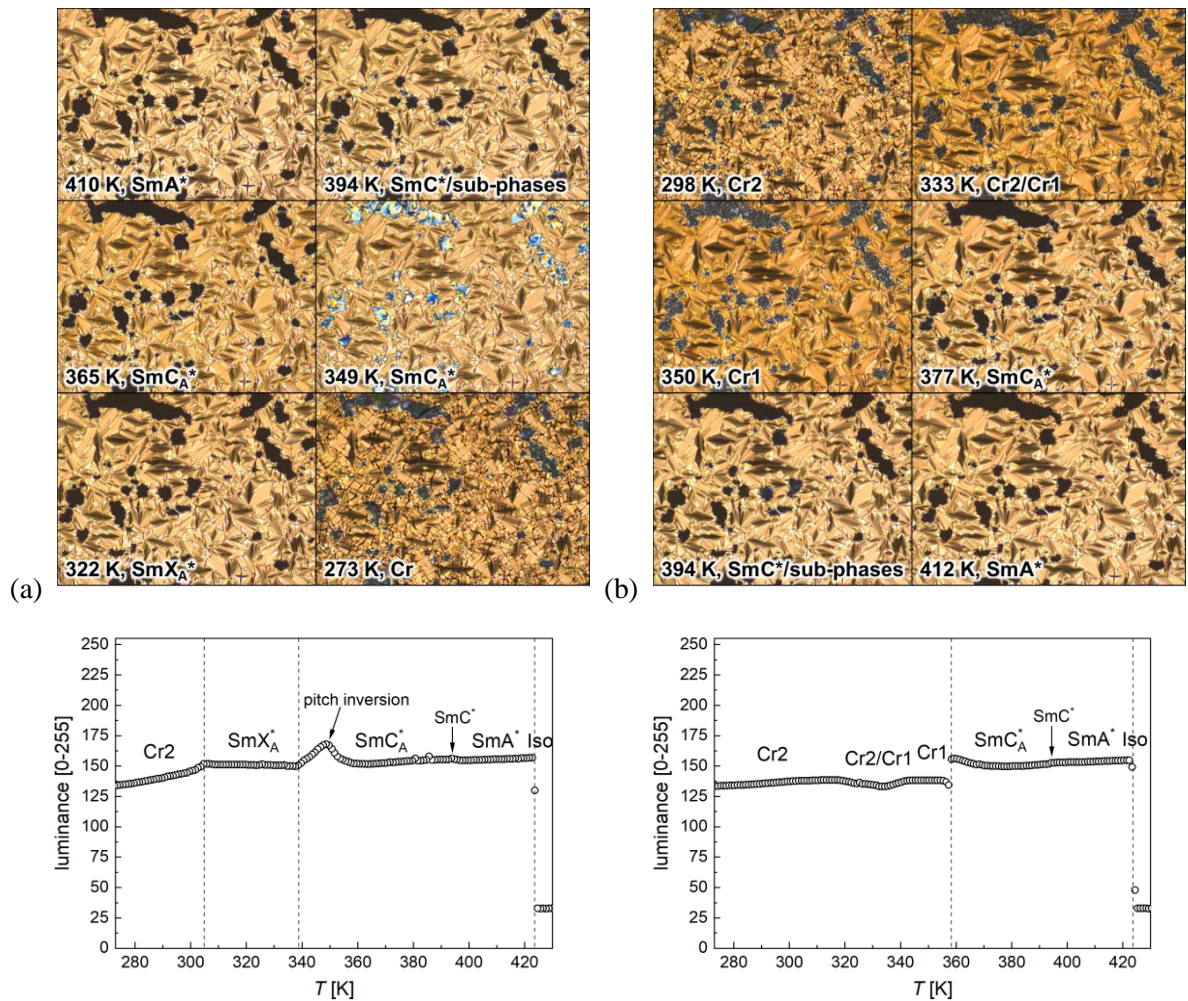

Figure S1. Selected POM textures of MHPOBC and average luminance of all textures collected during cooling (a) and heating (b) at 5 K/min. Each texture covers an area of 1243 × 933 μm$^2$.

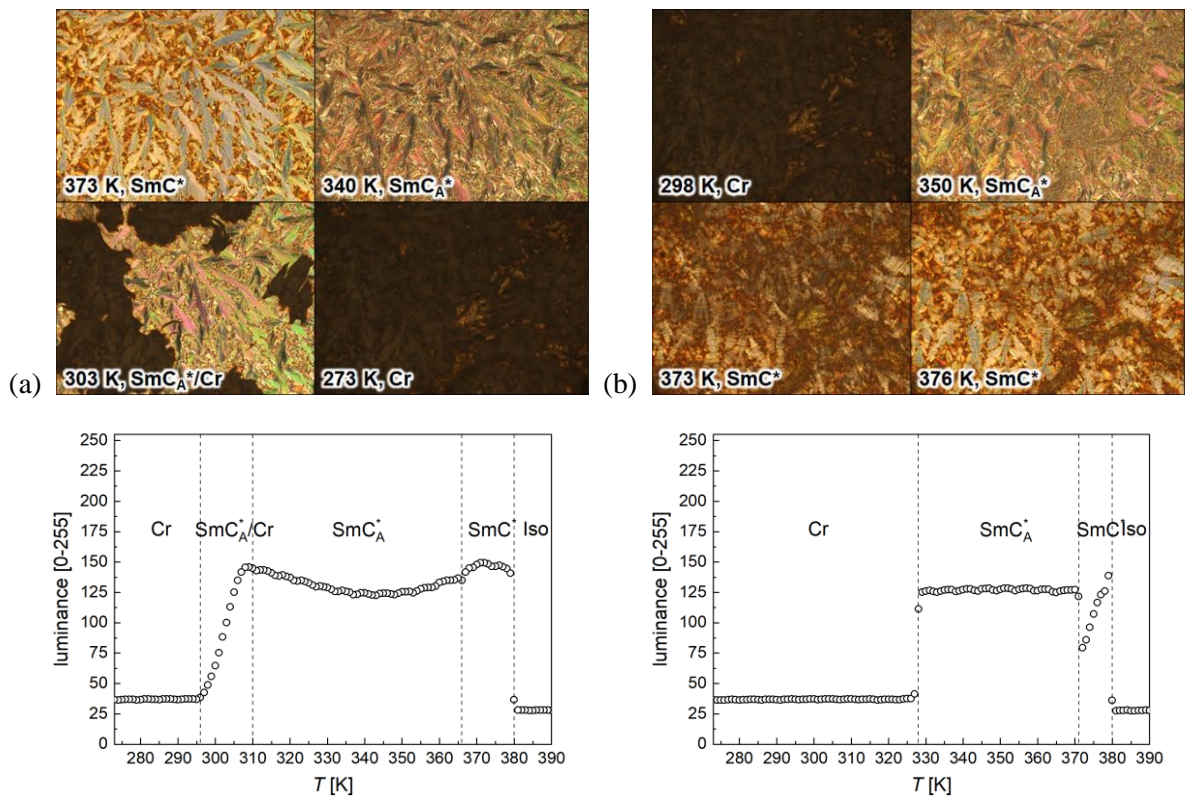

Figure S2. Selected POM textures of 3F2HPhF6 and average luminance of all textures collected during cooling (a) and heating (b) at 5 K/min. Each texture covers an area of 1243 × 933 μm$^2$.

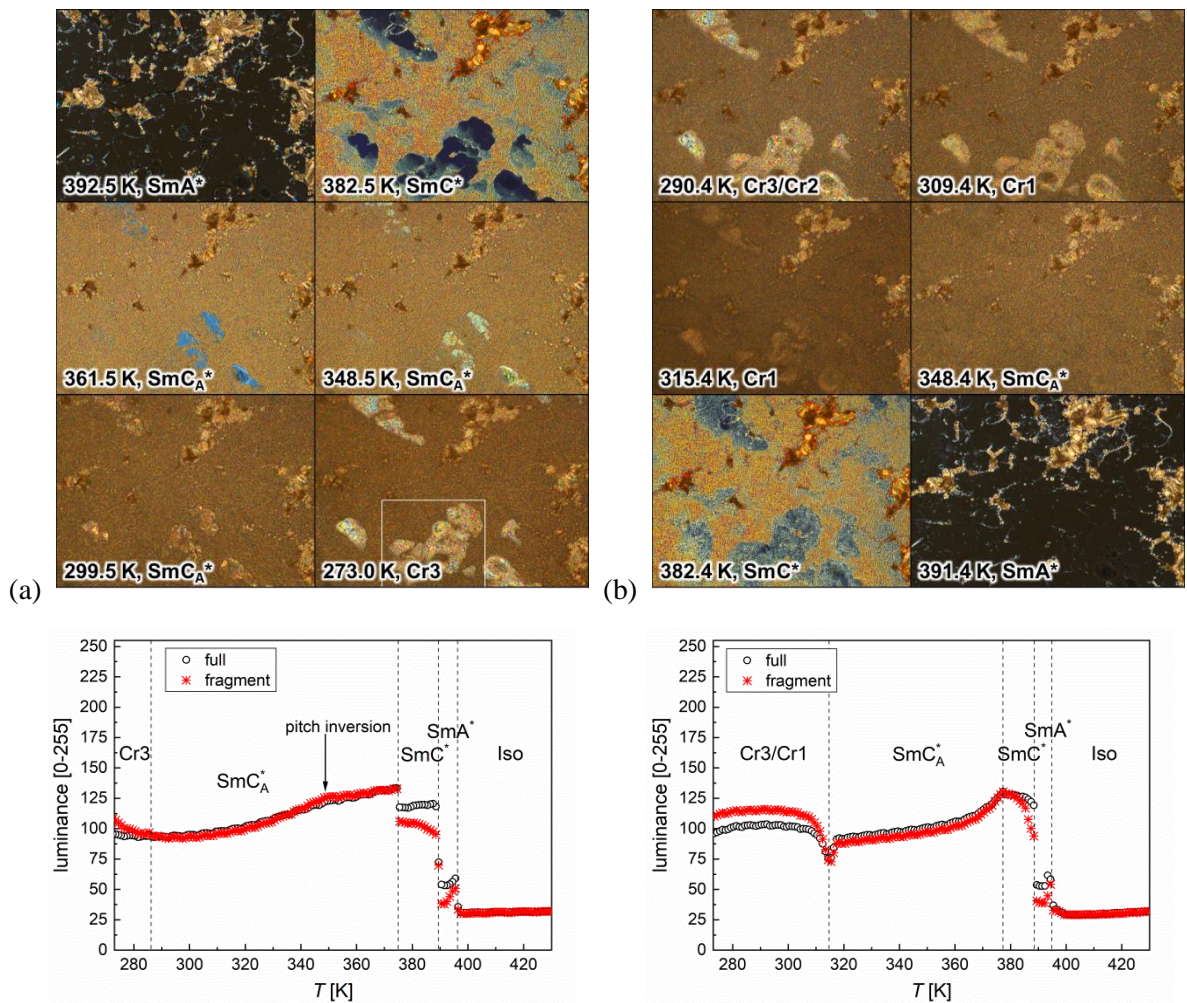

Figure S3. Selected POM textures of MIX2HF6 and average luminance of all textures collected during cooling (a) and heating (b) at 5 K/min. Each texture covers an area of 1243 × 933 μm$^2$.

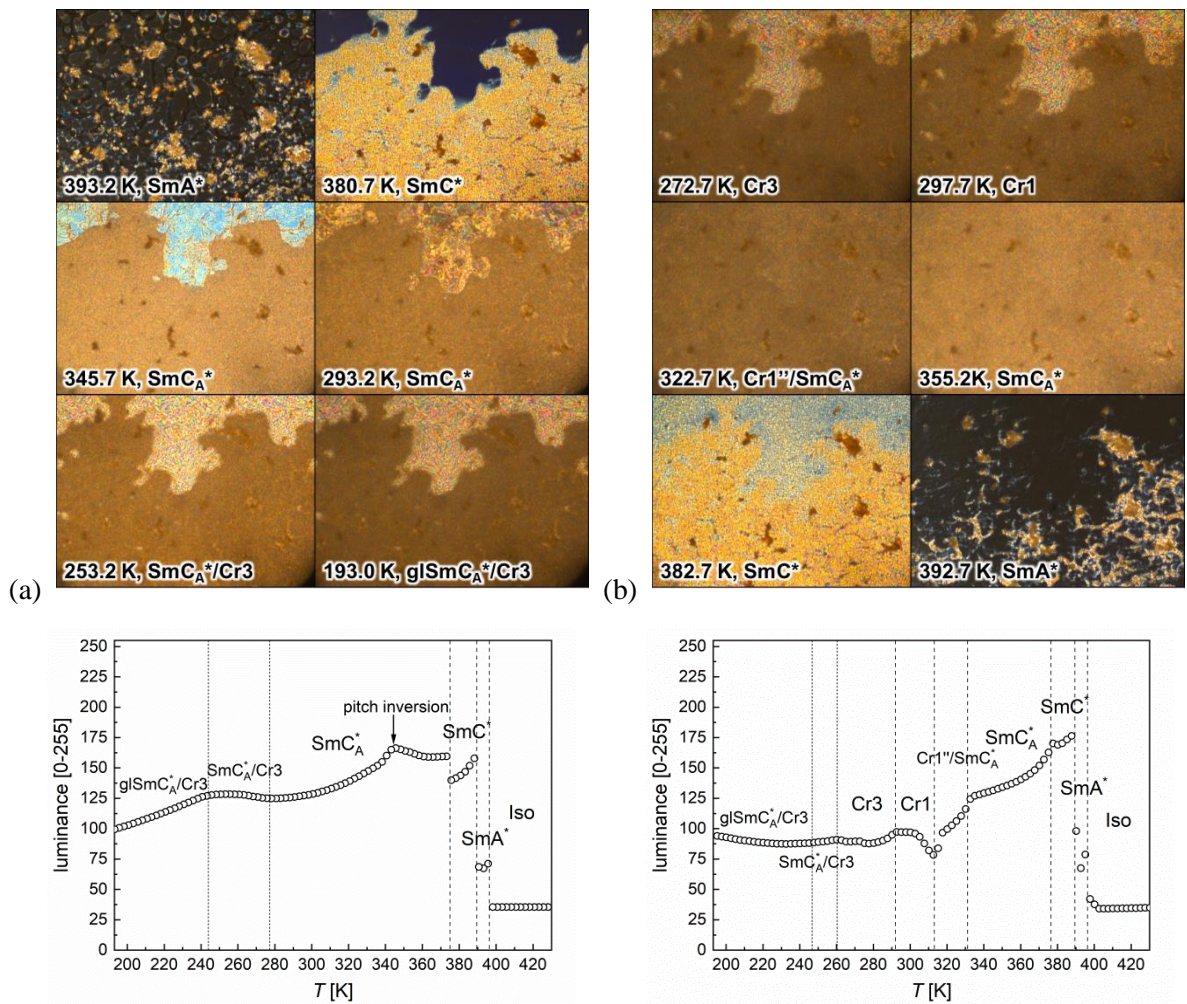

Figure S4. Selected POM textures of MIX2HF6 and average luminance of all textures collected during cooling (a) and heating (b) at 30 K/min. Each texture covers an area of 1243 × 933 μm². The lines indicating the crystallization and glass transition temperatures are based on DSC results for 30 K/min.